# ASSESSING THE IMPACT OF CONTACT TIME ON LEACHATE CHEMISTRY FROM RECYCLED CONCRETE AGGREGATES

**Morgan Sanger, Corresponding Author**
Department of Civil and Environmental Engineering, Geological Engineering Program
University of Wisconsin – Madison
660 N. Park Street, Madison, WI 53706
Email: msanger@wisc.edu

**Gabrielle Campagnola**
Department of Civil and Environmental Engineering
University of Wisconsin – Madison
660 N. Park Street, Madison, WI 53706

**Robin Ritchey**
Department of Civil and Environmental Engineering
University of Wisconsin – Madison
660 N. Park Street, Madison, WI 53706

**Tuncer Edil, Ph.D., P.E.**
Department of Civil and Environmental Engineering, Geological Engineering Program
University of Wisconsin – Madison
1415 Engineering Drive, Madison, WI 53706
Email: tbedil@wisc.edu

**Matthew Ginder-Vogel, Ph.D.**
Department of Civil and Environmental Engineering, Geological Engineering Program, and Environmental Chemistry and Technology Program
University of Wisconsin – Madison
660 N. Park Street, Madison, WI 53706
Tel: (608) 262 – 0768; Email: mgindervogel@wisc.edu

Word count: 6,196 words text + 4 tables/figures x 250 words (each) = 7,196 words (max 7,500)

Submission Date: August 1, 2018

**ABSTRACT**

Recycled concrete aggregate (RCA) is recognized as a readily available, mechanically sufficient construction and demolition waste product that is suitable as a base course substitute for natural, virgin aggregate in pavement construction. Environmentally responsible applications of RCA must consider the high alkalinity, high pH leachate, and heavy metal leaching risks reported in the literature. The existing body of literature does not address discrepancies between field and laboratory measurements of RCA leachate pH, nor are there any existing studies of aged RCA leachate composition as a function of time. To consider the influence of contact time on RCA leachate, the present study evaluates recovered RCA base course samples from the Minnesota Road Research highway construction study site using modified batch test methodology. Leachate pH, alkalinity, and calcium ion ($Ca^{2+}$) concentration were monitored for 24 hours to understand RCA leachate chemistry during the initial contact period. Leachate pH is high upon initial contact with water (pH > 10) and decreases over time as it reacts with atmospheric carbon dioxide. Calcium ion concentration increases rapidly in the initial contact period, then more gradually as calcium saturation is reached. Alkalinity stabilizes (50-65 mg $CaCO_3$/L) after a dramatic increase during the initial contact period.

## INTRODUCTION

The use of recycled concrete aggregate (RCA) as a substitute for natural, virgin aggregate in pavement construction applications is well established and successful, particularly as granular or stabilized base course. RCA exhibits desirable mechanical proprieties for pavement base course applications including high resilient modulus, low compacted unit weight, and freeze-thaw resistance (1, 2). Costs associated with the use of natural aggregates are largely incurred in transporting material from the quarry to the construction site. The use of RCA can be economically beneficial by eliminating costs related to mining and transporting materials (3). Similarly, the use of RCA exhibits life-cycle environmental benefits in conserving finite natural aggregate resources and reducing energy consumption, water usage, and carbon dioxide emissions associated with mining and transportation of aggregate material (4).

Recognized as a readily available, mechanically sufficient, construction and demolition waste product with life-cycle economic and environmental benefits, RCA is a wise choice of engineering material in pavement construction (1, 4–7). However, RCA base course construction should account for drain clogging and permeability loss in drainage systems caused by calcium carbonate precipitation out of RCA leachate (8, 9). Environmentally safe and responsible applications of RCA must also consider the high alkalinity and high pH leachate and heavy metal leaching risks that are reported in literature (10–17). Crushing concrete for use as RCA exposes fresh surfaces to atmospheric conditions, most notably carbon dioxide and rainwater infiltration. The fresh surfaces contain cement phases including Aft/Afm, calcium hydroxide, and calcium silicate hydrate (C-S-H) which can react with rain water to form high alkalinity and high pH leachate (18–24). RCA leachate generation from stockpiles and road base is unavoidable, and therefore it is of great interest to understand the fate and transport of the leachate, and whether pre-treatment, prescribed aging, or remediation is necessary to limit the environmental impact of RCA.

Several laboratory and field studies have investigated the environmental implications of RCA leachate chemistry in pavement base course applications. Much of the work to investigate RCA leachate chemistry pertains to the risk of trace element leaching (10, 13, 29–34, 14–17, 25–28). Long-term highway field studies of RCA leachate illustrate that, despite an initial elevated leachate pH (pH > 10), leachate pH approaches neutral pH approximately within a year of construction (11, 13, 16, 17). To investigate the pH and alkalinity of RCA leachate observed in the field studies, laboratory investigations have employed batch reactor leach tests (10, 14, 15), box leach tests (35–37), and column leach tests (28, 38); however, the persistently high leachate pH measured in the laboratory is in poor agreement with the near-neutral leachate pH measured in the field. It is likely that laboratory methods employed in the previous investigations may not be environmentally relevant for RCA application as pavement subbase.

There are many parameters that can change the surface chemistry of RCA in the field that are difficult to represent in laboratory studies. These include the frequency and duration of precipitation, degree of saturation, temperature, variation in subbase soil geology, and traffic loads, all of which vary in time and space. Depending on how well the RCA base course layer drains, water may be in contact with the RCA for as little as one or two hours, or longer than several days (39). Therefore, it is critical to understand how leachate chemistry changes as a function of time in order to predict possible environmental issues. The discrepancies in leachate pH reported in laboratory and field studies suggest that the material pH of RCA and subgrade soils determined from laboratory tests are not representative of the leachate pH that can be expected from field conditions (11, 13, 16).

The purpose of this study is to examine the leachate chemistry of field-deployed RCA recovered from the Minnesota Road Research (MnROAD) highway construction study site. The existing body of literature does not address the discrepancies between field and laboratory measurements of RCA leachate pH, and there are no existing studies of aged RCA leachate composition as a function of time. The present study evaluates the impact of contact time on RCA chemistry, and addresses the discrepancies between field and laboratory measurements of RCA leachate pH using modified batch reactor methodologies. Leachate pH, alkalinity, and calcium ion ($Ca^{2+}$) concentration were monitored for 24 hours to understand RCA leachate chemistry during the initial contact period.

## MATERIALS AND METHODS

**MnROAD Field Site Description**

In previous work, RCA material from Minnesota Department of Transportation was obtained during construction of roadway cells at the MnROAD test facility in Maplewood, Minnesota (*11, 28*). The site was constructed in September 2008 on the MnROAD test facility mainline westbound of I-94 between St. Cloud and Minneapolis, Minnesota. Experimental Cell 16 was constructed with Minnesota RCA base course aggregate to investigate the effects of RCA as base course on leachate pH and alkalinity (*40*). Pan lysimeters (3 m x 3 m) were installed under each of the test cells to collect percolated leachate. Additional site construction details are described by Chen et al. (2013) (*11*).

In the initial time period after construction, leachate collected from the pan lysimeters measured between pH 7.5 and 8.5 (*11*). Leachate pH measured between 7.2 and 7.4 in samples collected from April 2016 to July 2016, approximately eight years after construction (*40*).

The test sections were deconstructed in July 2016 and aggregate samples were collected from Cell 16. During deconstruction, RCA material was sampled from the base course layer below the passing lane, driving lane, and centerline of the MnROAD research facility: samples 16P-1, 16D-1, and 16C-1, respectively (*40*). The RCA material analyzed in this study are the recovered 16P-1, 16D-1, and 16C-1 RCA samples.

The physical and hydraulic properties of the original MnROAD RCA materials were characterized prior to installation using grain size analysis, residual water content, dry unit weight, absorption, specific gravity, and compaction characteristics (*11, 28*). The recovered MnROAD RCA materials were characterized using the same analyses after eight years of field deployment (*40*). No significant differences in physical and hydraulic properties were observed when comparing the original and the recovered MnROAD RCA; however, the carbonate content of the recovered RCA was higher than that of the original RCA material, indicating carbonation during the eight-year field deployment (*40*).

**Batch Reactor Methodology**

The MnROAD RCA base course samples were homogenized by hand mixing and oven-dried overnight. Each base course sample was tested in triplicate using batch reactors prepared with an initial solid to liquid ratio of 1:10, 50 g of base course to 500 mL Milli-Q Integral Ultrapure Water (MQ) (18.2 MΩ·cm). Batch reactors were constructed as continuous stirred tank reactors and exposed to atmospheric carbon dioxide to simulate environmentally relevant, open system conditions. Guidance for developing batch reactor test methodology is provided by Kosson et al. (2002), where test method recommendation SR002.1 – Alkalinity, Solubility, and Release as a function of pH recommends batch reactors with liquid to solid ratio of 10, and the use of agitation throughout the duration of the experiment via an end-over-end tumbler at 28 rpm (±2 rpm) (*41*).

Particle abrasion from vigorous agitation (e.g., and end-over-end tumbler) results in degradation of surface coatings (*42*). With time, intermittent saturation, and exposure to atmospheric carbon dioxide, a calcium carbonate surface coating precipitates on the surface of RCA, called carbonation, which prevents dissolution of minerals contributing to high pH leachate (*43*). In the case of previous studies, the utilization of an end-over-end tumbler to determine material pH likely caused particle abrasion and removal of the protective calcium carbonate surface coating that would otherwise limits the leachate pH by reducing the dissolution of minerals contributing to high pH leachate. Vigorous shaking and particle abrasion are not relevant to RCA base course in the field, thus any surface coating that forms as a result of time, intermittent saturation, and exposure to atmospheric carbon dioxide is assumed to remain intact. To ensure homogeneity while minimizing abrasion between particles, the batch test method used was modified to exclude the use of an end-over-end tumbler and instead a shaker plate was utilized to continuously, gently shake the batch reactor for 24 hours. to more effectively simulate field conditions for RCA pavement base course applications.

Calcium ion concentration and pH of the RCA leachate in the batch reactor were measured using the Thermo Scientific Orion Combination Calcium Electrode and the Thermo Scientific Orion Combination pH Electrode, respectively. Resolution of the $Ca^{2+}$ concentration was limited by the precision of the Thermo Scientific Orion Combination Calcium Electrode to 0.0001 M. The elemental composition of MnROAD

RCA leachate was previously quantified; the $Ca^{2+}$ concentration was found to be an order of magnitude greater than Na and K and three orders of magnitude greater than Cr and Zn (*40*). Consequently, interferences from other ions resulted in less than ± 10% uncertainty in measured $Ca^{2+}$ concentrations.

To determine alkalinity at different time points, a 6-mL leachate sample from the batch reactor was filtered using Millipore 0.2-μm Isopore Membrane Filters and diluted with 34 mL of MQ, then titrated in a Mettler Toledo Compact Titrator to pH 4.5 with 0.01 N $H_2SO_4$. Because the greatest variation in RCA leachate chemistry occurs in the initial period of contact between the RCA base material and water, the sample frequency is higher at the beginning of the experiment. Synchronous measurements of pH, $Ca^{2+}$ concentration, and alkalinity samples were taken at the following time points in minutes from initial contact time: 1, 2, 3, 4, 5, 6, 7, 8, 9, 10, 15, 20, 25, 30, 40, 50, 60, 90, 120, 180, 240, 300, 360, 1440 minutes.

## RESULTS AND DISCUSSION

RCA leachate pH (Figure 1), $Ca^{2+}$ concentration (Figure 2), and alkalinity (Figure 3) are monitored for 24 hours in a batch reactor. The initial hour is provided with greater resolution to illustrate the significant changes observed in this period in RCA leachate pH (Figure 4), $Ca^{2+}$ concentration (Figure 5), and alkalinity (Figure 6). The data points represent the mean result for the triplicate batch reactors with a range bar to illustrate the maximum and minimum value observed. The results are discussed in terms of leachate chemistry and in terms of the different sampling locations beneath the MnROAD test facility. RCA leachate pH is also discussed in comparison the bulk material pH determined previously by the authors.

## RCA Leachate Chemistry

The present study focuses on RCA leachate chemistry in terms of leachate pH, $Ca^{2+}$ concentration, and alkalinity. pH is defined as the negative log of the activity of hydrogen ions ($H^+$) in solution (Equation 1). In many freshwater systems, and in the case of the present study, the activity can be considered to be the concentration $H^+$. Neutral water has both hydroxide ($OH^-$), a strong base, and hydrogen ions, a strong acid, present in equal amounts (Reaction 2). Water in an open system with atmospheric carbon dioxide will dissolve gaseous carbon dioxide ($CO_2$) from the atmosphere to form aqueous carbon dioxide (Reaction 3) (*44*). Aqueous carbon dioxide reacts with the water to form carbonic acid ($H_2CO_3$), (Reaction 4). Carbonic acid reacts with hydroxide to form bicarbonate ($HCO_3^-$) (Reaction 5), and bicarbonate then reacts with hydroxide to form carbonate ($CO_3^{2-}$), (Reaction 6). Alkalinity is defined as the acid neutralization capacity of a solution, also known as the buffering capacity. Alkalinity is equal to the sum of the concentration of strong and weak bases in solution minus the concentration of strong acids (Equation 7). Since both bicarbonate and carbonate are weak bases, the consumption of hydroxide to form bicarbonate and carbonate decreases the pH of the leachate but does not decrease the alkalinity.

Portland cement concrete, which is crushed after its usable life as a monolith to become RCA, is a mixture of coarse aggregate in Portland cement paste consisting of calcium carbonate ($CaCO_3$), Afm/Aft, calcium hydroxide, known as portlandite, ($Ca(OH)_2$), and calcium-silicate hydrate (C-S-H) ($3CaO \cdot 2SiO_2 \cdot 3H_2O$) (*14, 43, 45–49*). When saturated with water, RCA interacts chemically with the carbonate system of natural waters. In laboratory batch reactors and in pavement base course applications exposed to meteoric water, water dissolves portlandite and calcium carbonate on RCA surfaces (Reactions 8 and 9, respectively). While dissolution and carbonation of AFm/Aft may contribute to the dissolved $Ca^{2+}$ concentration, $Ca2^+$ is largely supplied by the dissolution of calcium carbonate and portlandite. Similarly, AFm/Aft carbonation and dissolution may contribute leachate pH, dissolution of portlandite, which adds additional hydroxide to the system, thus increasing the pH, is required to achieve the high pH observed in RCA leachate chemistry.

The total dissolved carbonate ($H_2CO_3^*$, $HCO_3^-$, $CO_3^{2-}$) is product of the reactions involving carbon dioxide gas ($CO_2(g)$) dissolution into the aqueous phase. The rate of which will be controlled by the amount of portlandite and calcium carbonate that dissolves into the solution. Carbonate combines with $Ca^{2+}$ in solution and precipitates calcium carbonate. With time and exposure to carbon dioxide in the atmosphere, both portlandite and C-S-H react with aqueous carbon dioxide to precipitate calcium carbonate (Reactions 10 and 11, respectively) (*43, 45*).

The formation of calcium carbonate on the surface of RCA is referred to as carbonation, which can limit further dissolution of calcium hydroxide, therefore mitigating the source of high pH leachate. Carbonation is the formation of a calcium carbonate layer on the RCA surface by Reactions 9, 10, and 11 (*49*). The calcium carbonate layer inhibits the further dissolution of cement hydrate phases and therefore, carbonation is a desirable property of RCA as it limits high pH of the leachate (*17, 50, 51*). Carbonation requires a thin film of water on the RCA surface to supply aqueous carbon dioxide for reaction, and completely saturated RCA will take longer to carbonate as the process is limited by the diffusion of atmospheric carbon dioxide (*49, 50, 52*).

$$pH = -\log{[H^+]} \tag{1}$$
$$2H^+ + 2OH^- \rightarrow 2H_2O \tag{2}$$
$$CO_2(g) \leftrightarrow CO_2(aq) \tag{3}$$
$$H_2O + CO_2(aq) \rightarrow H_2CO_3^* \tag{4}$$
$$H_2CO_3^* + OH^- \rightarrow H_2O + HCO_3^- \tag{5}$$
$$HCO_3^- + OH^- \rightarrow CO_3^{2-} + H_2O \tag{6}$$
$$Alkalinity = \sum[strong\ bases] + \sum[weak\ bases] - \sum[strong\ acids] \tag{7}$$
$$Ca(OH)_2 \rightarrow Ca^{2+} + 2OH^- \tag{8}$$
$$CaCO_3 \leftrightarrow Ca^{2+} + CO_3^{2-} \tag{9}$$
$$Ca(OH)_2 + CO_2(aq) \rightarrow CaCO_3 + H_2O \tag{10}$$
$$(3CaO \cdot 2SiO_2 + 3CO_2(aq) \rightarrow (3CaCO_3 \cdot 2SiO_2 \cdot H_2O) \tag{11}$$

## Impact of Contact Time on Recovered RCA Leachate Chemistry

The chemical behavior of RCA leachate observed in the present laboratory batch reactors illustrates the expected chemical behavior. The initial high pH (pH > 10) is the result of additional hydroxide ions to the leachate introduced via portlandite dissolution (Reaction 8) (Figures 1 and 4). Leachate pH decreases steadily over the duration of the experiment as hydroxide, a strong base, is consumed to produce bicarbonate and carbonate, both weak bases (Reactions 5 and 6) (Figure 1). Reduction of leachate pH is thereby influenced by the rate of atmospheric carbon dioxide dissolution into the leachate (Reaction 3) to form carbonic acid (Reaction 4) to react with the hydroxide and/or C-S-H (Reactions 5 and 6) (*44*).

The high pH of RCA leachate (pH > 10), which is one primary environmental concern associated with the use of RCA base course, was observed within the initial hours of the experiment as a result of portlandite dissolution (Reaction 8) (Figure 1). After the initial peak, pH decreased steadily for the remainder of the 24-hour experiment as hydroxide is consumed to produce bicarbonate and carbonate, both weak bases (Reactions 5 and 6) (Figure 1). Reduction of leachate pH is thereby influenced by the rate of atmospheric carbon dioxide dissolution into the leachate (Reaction 3) to form carbonic acid (Reaction 4) to react with the hydroxide and/or C-S-H (Reactions 5 and 6) (*44*). The decrease in leachate pH throughout the duration of the experiment is expected to continue asymptotically towards the near-neutral pH (pH 7.5-8.5) measured at the MnROAD field in the initial period after construction (*11, 28*). In addition to neutralization via carbon dioxide infiltration, it is expected that soil minerals in the subgrade will also react with the hydroxide, thus neutralization will occur more quickly in field applications as the leachate drains from the RCA base course layer into the subgrade and interacts with soil (*44*).

Calcium ion concentration increases rapidly in the first hours of the experiment as portlandite and calcium carbonate on the surfaces of the RCA dissolve (Reactions 8 and 9) (Figure 2). Calcium ion concentration increases more gradually thereafter as concentration approaches saturation and as $Ca^{2+}$ is removed from the system by calcium carbonate precipitation (Reaction 9) (Figure 2).

Alkalinity increases rapidly in the first hours of the experiment as the dissolution of portlandite introduces hydroxide, a strong base (Reaction 6) (Figure 3). Alkalinity then stabilizes after this initial increase because the reactions of strong base to form weak bases (Reactions 5 and 6) maintain the total amount of base in the system (Figure 3). Decreases in alkalinity, as observed in Figures 3a and 3b, result from the consumption of carbonate by calcium carbonate precipitation (Reaction 9).

**Spatial Variation in RCA Leachate Chemistry**

Observed leachate chemistry is slightly different for samples obtained from different locations beneath the roadway with respect to pH. RCA material obtained from the centerline of the road exhibits the lowest initial and final pH values. The centerline leachate pH begins at a maximum of 10.2 and ends at a minimum 9.4 (Figure 1c). Material obtained from the passing and driving lanes exhibit similar pH trends in the 24-hour testing period. The passing lane leachate pH peaks at pH 10.6 and decreases to a minimum pH of 9.7 (Figure 1a), and the driving lane leachate pH begins at a maximum of 10.5 and ends at a minimum 9.9 (Figure 1b). The maximum pH value is measured within the first hour for all three samples. The small but consistent differences in leachate pH from the different samples is likely due to different levels of carbonation of the RCA material from the locations and fractional differences in portlandite content (*40*).

Calcium ion concentration trends are similar for material from the centerline, passing, and driving lanes. The passing lane leachate has an initial $Ca^{2+}$ concentration of 0.0002 M and a 0.0017 M final concentration (Figure 2a), the driving lane leachate has an initial $Ca^{2+}$ concentration of 0.0003 M and a 0.0017 M final concentration (Figure 2b), and the centerline leachate has an initial $Ca^{2+}$ concentration of 0.0001 M and a 0.0018 M a final concentration (Figure 2c). Most of the $Ca^{2+}$ ions are released within the first 4 to 6 hours of the batch reactor experiments.

Alkalinity of the leachate from the passing lane exhibits an initial increase from 36.4 mg $CaCO_3$/L to a peak value of 61.7 mg $CaCO_3$/L, followed by a decrease and stabilization around 56.2 $CaCO_3$/L (Figure 3a). Similarly, alkalinity of the leachate from the driving lane exhibits a rapid initial increase from 41.8 mg $CaCO_3$/L to a peak value of 58.5 mg $CaCO_3$/L, and a consequent rapid decrease and stabilization and 48.8 mg $CaCO_3$/L (Figure 3b). Alkalinity of the leachate from the centerline exhibits a rapid initial increase from 32.3 mg $CaCO_3$/L and appears to peak stabilize around 61.8 mg $CaCO_3$/L (Figure 3c). The primary sources of alkalinity in RCA leachate are hydroxide, carbonate, and bicarbonate resulting from the dissolution of portlandite and calcium carbonate (Reactions 8 and 9). Additional sources of alkalinity, including dissolved silicate and phosphate, are negligible in comparison to hydroxide, bicarbonate, and carbonate (*11, 28*).

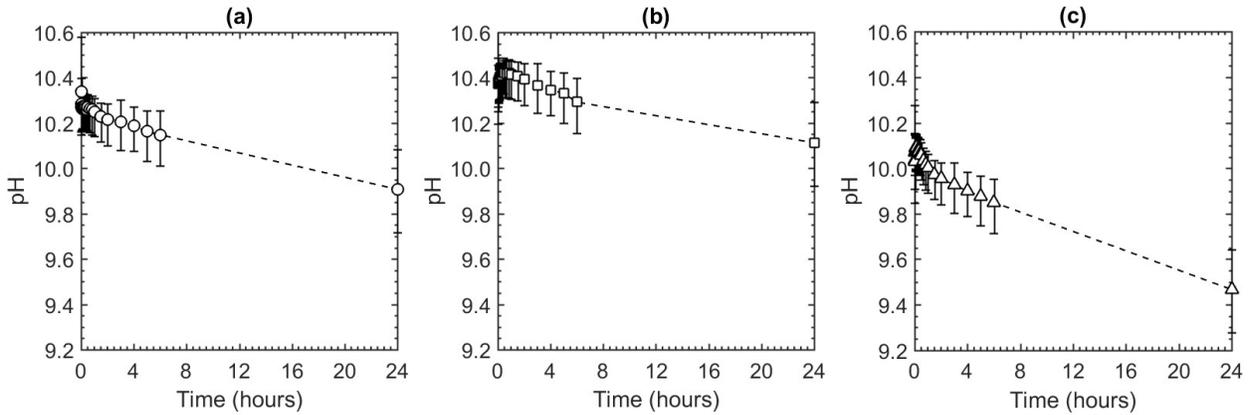

**FIGURE 1** pH of RCA leachate measured during the 24-hour modified batch test for RCA leachate samples (a) 16P-1; (b) 16D-1; (c) 16C-1.

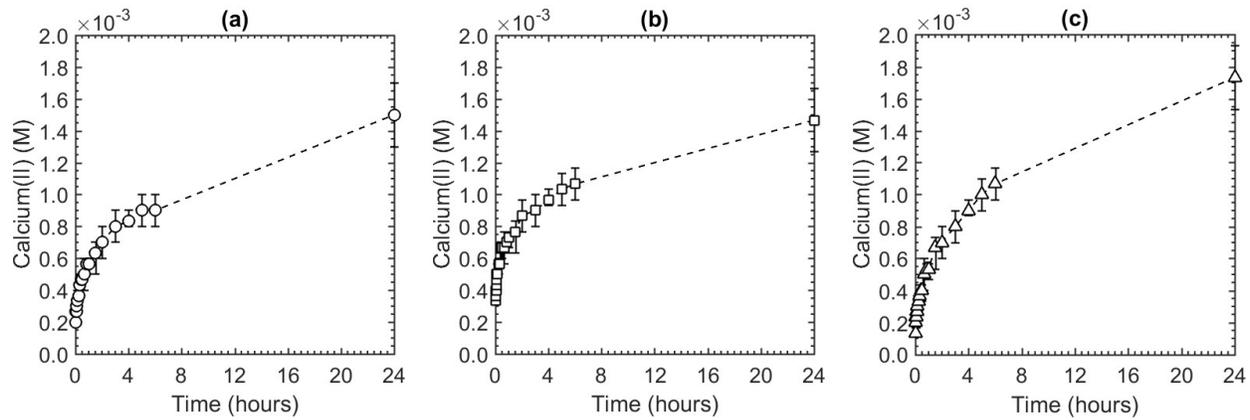

**FIGURE 2** Calcium ion concentration of RCA leachate measured during the 24-hour modified batch test for RCA leachate samples (a) 16P-1; (b) 16D-1; (c) 16C-1.

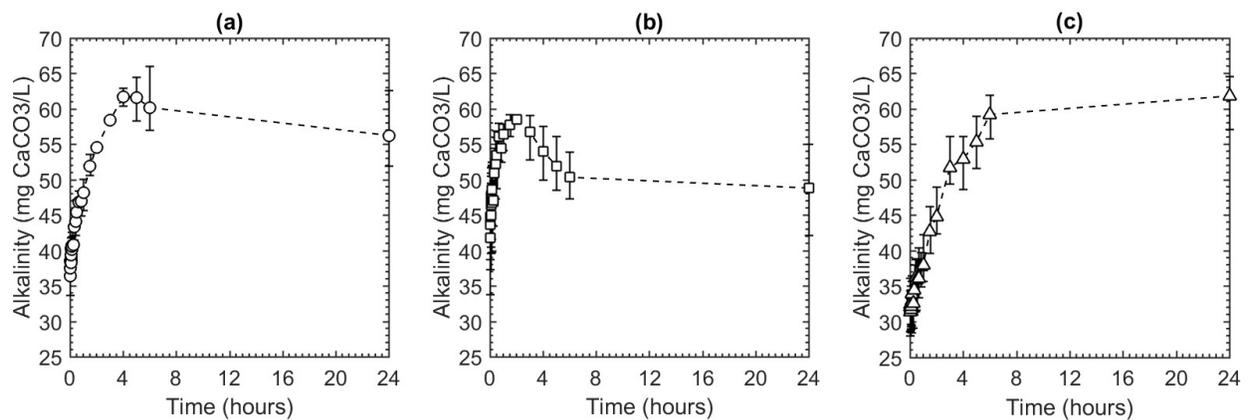

**FIGURE 3** Alkalinity of RCA leachate measured during the 24-hour modified batch test for RCA leachate samples (a) 16P-1; (b) 16D-1; (c) 16C-1.

**Environmental Relevance of Modified Batch Reactors**

The material pH of the recovered MnROAD RCA samples was determined by Natarajan et al. using an end-over-end tumbler, as is standard in conventional batch tests (*40*, *41*). To determine material pH for the recovered MnROAD RCA samples in the present study, conventional batch test methodology is modified to use a shaker plate instead of an end-over-end tumbler, keeping all else constant in the procedure. In both the conventional and modified batch test methodologies, pH of the RCA leachate is measured after 24 hours of contact time in the respective batch reactors, called the material pH. The material pH of the RCA leachate determined in the present study by modified batch tests is lower than that determined by conventional batch tests methods (Table 1).

The particle abrasion that occurs in an end-over-end tumbler is not environmentally relevant for pavement systems in which base course aggregate does not move significantly once constructed. By using a shaker plate instead of an end-over-end tumbler, particle abrasion is limited, and the protective calcium carbonate surface coating is preserved. Preservation of the calcium carbonate surface coating limits the dissolution of cement hydrate phases, such as portlandite, which are responsible for increasing the leachate pH. Thus, the use of a shaker plate in the modified batch reactors of the present study simulates a more environmentally representative conditions and yields pH values more relevant to field application.

**TABLE 1 Bulk material pH determined previously using conventional batch test methods by Natarajan et. al (*40*) and leachate pH determined using modified batch test methods.**

|  | 16 P | 16 D | 16 C |
|---|---|---|---|
| **Material pH – Conventional Method** | 11.3 | 11.4 | 11.1 |
| **Material pH – Modified Method** | 9.9 | 10.1 | 9.5 |

The environmental relevance of the modified batch reactors is further demonstrated in the decrease in pH measured with contact time. Previous laboratory investigations of MnROAD RCA materials exhibited persistently high pH over several pore-volumes (*11*, *28*). These persistently high pH measurements did not correlate with to the near-neutral leachate pH measured at the MnROAD field site (*11*, *28*). Leachate pH degradation measured in the present 24-hour laboratory investigations are consistent with field measurements and therefore confirm that RCA in base course applications was be environmentally safe and responsible use of RCA in pavement base course applications.

**CONCLUSIONS AND RECOMMENDATIONS**

The present assessment of RCA leachate chemistry demonstrates pH, $Ca^{2+}$ concentration, and alkalinity values consistent with the chemical behavior expected of the system. Leachate pH is high upon initial contact with water (pH > 10) and decreases over time as it reacts with carbon dioxide. In addition to neutralization via carbon dioxide infiltration, it is expected that soil minerals in the subgrade will also react with the hydroxide, thus pH degradation will occur more quickly in field applications as the leachate drains from the RCA base course layer into the subgrade and interacts with soil (*44*). Calcium ion concentration increases rapidly in the initial hours of the experiment as calcium carbonate and portlandite dissolve, then more gradually as the concentration of the dissolved ions approaches saturation and calcium carbonate begins to precipitate. Alkalinity stabilizes between 50-65 mg $CaCO_3$/L after a rapid initial increase following the dissolution of portlandite and calcium carbonate.

RCA chemistry in a laboratory setting is evaluated using a shaker plate instead of an end-over-end tumbler to minimize the impact of particle abrasion and improve environmental relevance. The pH of RCA leachate determined in the present study by modified batch reactors is lower than that previously determined by conventional batch tests methods (*40*). The results demonstrate that particle abrasion that occurs in an end-over-end tumbler is not environmentally relevant for pavement systems, and that the use of a shaker plate in the present study provides more environmentally-relevant results.

The present study also addresses the discrepancies between field and laboratory measurements of RCA leachate pH in pavement base course applications. Field measurements of MnROAD RCA leachate pH measured leachate pH to be between 7.5 and 8.5, close to neutral, in the initial period after construction at the MnROAD field site (*11*, *28*). The near-neutral leachate pH measured in the field had not previously been replicated in laboratory investigations. The decrease in leachate pH throughout the duration of the experiment is expected to continue asymptotically towards the near-neutral field pH measurements.

The work presented in this study is the initial step to understand the generation, fate and transport of RCA leachate, and whether pre-treatment, prescribed aging, or remediation is necessary to limit the environmental impact of RCA leachate. Additional work is required to characterize the leachate chemistry of freshly-crushed RCA and utilize geochemical modelling to connect the observed differences in RCA properties (e.g., grain size distribution, solid phase composition) to leachate to differences in leachate chemistry. Additional work is also required to monitor leachate chemistry after phase separation to simulate drainage from the base course layer into the subgrade and hydrogeologic system. Integration of laboratory chemistry and geochemical modelling will then inform the development of industry guidelines for prescribed aging or stockpiling criteria for the safe and responsible use of RCA as pavement base course.

## ACKNOWLEDGMENTS

This work was funded by the Recycled Materials Resource Center by a pooled fund of eight member states (IA, IL, MN, NC, PA, VA, WA, WI). The Recycled Materials Resource Center is supported through the Federal Highway Administration. Funding was also provided by the Portland Cement Association and the Ready Mixed Concrete Research and Education Foundation. The opinions, findings, conclusions, and recommendations expressed herein are those of the authors and do not necessarily represent the views of the sponsors.

## AUTHOR CONTRIBUTION STATEMENT

The authors confirm contribution to the paper as follows: study conception and design: M. Ginder-Vogel, M. Sanger, T. Edil; data collection: G. Campagnola, R. Ritchey, M. Sanger; analysis and interpretation of results: M. Sanger, M. Ginder-Vogel, T. Edil; draft manuscript preparation: M, Sanger. All authors reviewed the results and approved the final version of the manuscript.